\documentstyle[sprocl,epsfig]{article}

\newcommand{\ol}{\overline}
\newcommand{\Pslash}{\kern 0.2 em P\kern -0.56em \raisebox{0.3ex}{/}}
\newcommand{\pslash}{\kern 0.2 em p\kern -0.4em /}
\newcommand{\kslash}{\kern 0.2 em k\kern -0.45em /}
\newcommand{\Sslash}{\kern 0.2 em S\kern -0.56em \raisebox{0.3ex}{/}}
\newcommand{\Mslash}{\kern 0.2 em M\kern -0.70em \raisebox{0.3ex}{/}}
\newcommand{\g}{\gamma}
\newcommand{\sig}{\sigma}
\newcommand{\eps}{\epsilon}
\newcommand{\dg}{\dagger}
\newcommand{\one}{1\hspace{-2pt}\rule[0.08ex]{0.45pt}{1.5ex}\hspace{2pt}}

\newcommand{\sT}{{\scriptscriptstyle T}}
\newcommand{\nn}{\nonumber}
\newcommand{\newangle}{{<\kern -0.3 em{\scriptscriptstyle )}}}

\begin{document}
\title{TWO-HADRON INCLUSIVE DIS AND INTERFERENCE FRAGMENTATION FUNCTIONS}

\author{M.~Radici}
\address{Dipartimento di Fisica Nucleare e Teorica, Universit\`{a} di 
Pavia, and\\
Istituto Nazionale di Fisica Nucleare, Sezione di Pavia, 
I-27100 Pavia, Italy}

\maketitle

\abstracts{
We investigate the properties of interference fragmentation functions 
arising from the emission of two leading hadrons inside the same jet for 
inclusive lepton-nucleon deep-inelastic scattering. Using an extended 
spectator model we give numerical estimates for the example of
the fragmentation into a proton-pion pair with its invariant 
mass on the Roper resonance.}

\section{Introduction}\label{sec:intro}

For the investigation of the nonperturbative nature of quarks and gluons
inside hadrons we mainly rely on the information extracted from
distribution (DF) and fragmentation functions (FF) in hard scattering
processes. There are three fundamental quark DF that completely 
characterize the quark inside hadrons at leading twist with respect to 
its longitudinal momentum and spin: the momentum distribution $f_1$, the 
helicity distribution $g_1$ and the transversity distribution $h_1$. 
Despite the first two ones, $h_1$ is difficult to address because of its
chiral-odd nature. A complementary information can come from the analysis 
of the hadrons produced by the fragmentation process of the final quark, 
namely from FF. So far, only the leading unpolarized FF, $D_1$, is partly 
known, which is the counterpart of $f_1$. The basic reason for such a poor 
knowledge is related to the difficulty of measuring more exclusive 
channels in hard processes. However, a new generation of experiments 
(including both ongoing measurements like HERMES and future projects like 
COMPASS or RHIC) allow for a much more powerful final-state identification 
and, therefore, for a wider and deeper analysis of FF, particularly when 
Final State Interactions (FSI) are considered. In this context, naive 
``T-odd'' FF naturally arise because the existence of FSI prevents 
constraints from time-reversal invariance to be applied to the 
fragmentation process~\cite{vari1}. This new set of FF includes also 
chiral-odd objects that become the natural partner needed to isolate $h_1$.

The presence of FSI allows that in the fragmentation process there are at
least two competing channels interfering through a nonvanishing phase. 
However, as shown in the following, this is not enough to generate 
``T-odd'' FF. A genuine difference in the Lorentz structure of the 
vertices describing the fragmenting processes is needed. This poses a 
serious difficulty in modelling the quark fragmentation into one observed 
hadron because it requires the ability of modelling the FSI between the 
hadron itself and the rest of the jet, unless one accepts to give up the 
concept of factorization. Therefore, here we will consider a hard process, 
semi-inclusive Deep-Inelastic Scattering (DIS), where the hadronization 
leads to two observed hadrons inside the same jet. We will determine all 
possible FF at leading twist, analyzing their symmetry properties. For the 
hadron pair being a proton and a pion with invariant mass equal to the 
Roper resonance, we will estimate FF using an extended version of the 
diquark spectator model~\cite{spectator}. In this case, FSI come from the 
interference between the direct production of the two hadrons and the 
decay of the Roper resonance.

\section{Why a fragmentation into two hadrons?}\label{sec:two}

Let's consider the situation of a one-hadron semi-inclusive DIS. Excluding
{\it ab initio} any factorization breaking mechanism, there are basically
two ways to describe the residual interactions of the leading hadron
inside the jet: to model microscopically independent interaction vertices
that lead to interfering competing channels, or to assume the hadron
moving in an external effective potential. In the former case, the
difficulty consists in modelling a genuine interaction vertex that cannot
be effectively reabsorbed in the soft part describing the hadronization.
In the latter, introduction of an external potential in principle breaks
the translational and rotational invariance of the problem. Further
assumptions can be made about the symmetries of the potential, but at the
price of loosing interesting contributions to the amplitude such as those
coming from naive ``T-odd'' FF.

\begin{figure}[h]
\begin{center}
\psfig{file=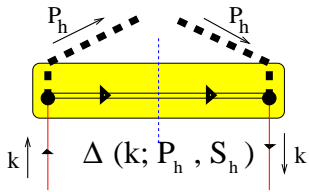, width=4cm}
\end{center}
\caption{Quark-quark correlation function $\Delta$ for the 
fragmentation of a quark into a hadron.}
\label{fig1}
\end{figure}

In fact, let's consider the pedagogical example where a quark with
momentum $k$ fragments into a leading hadron detected with momentum $P_h$
and mass $M_h$ (see Fig.~\ref{fig1}). The hadron is not polarized and does 
not interact with the rest of the jet, therefore its wave function is 
described by a free Dirac spinor $u(P_h)$. The jet itself is replaced by a 
spectator system which, again for sake of simplicity, is assumed to be a 
structureless on-shell scalar diquark with mass $M_D$ and momentum $k-P_h$ 
in order to preserve momentum conservation at the vertex. All this amounts 
to describe the remnants of the fragmentation process with a simple 
propagator line $\delta ((k - P_h)^2 - M_D^2)$ for the point-like on-shell 
scalar diquark $k-P_h$. Then, ignoring the inessential $\delta$ functions, 
the function $\Delta (k,P_h; S_h=0)$ describing the hadronization (the
soft blob in Fig.~\ref{fig1}) becomes
\begin{equation}
\Delta (k,P_h; S_h=0) \sim \displaystyle{\frac{-{\rm i}}{\kslash-m}} 
u(P_h)\overline{u}(P_h) \displaystyle{\frac{{\rm i}}{\kslash-m}}  
= \displaystyle{\frac{\kslash+m}{k^2-m^2}} (\Pslash_h+M_h) 
 \displaystyle{\frac{\kslash+m}{k^2-m^2}}  ,
\label{eq:1hsimple}
\end{equation}
where in the second line the usual projector for two free fermion spinors 
has been used. Eq. (\ref{eq:1hsimple}) can be cast in the following linear 
combination of all the independent Dirac structures of the process allowed 
by parity invariance~\cite{ralsop},  
\begin{equation}
\Delta (k,P_h; S_h=0) = A_1 M_h + A_2 \Pslash_h + A_3 \kslash + 
\displaystyle{\frac{A_4}{M_h}} \sigma_{\mu \nu} P_h^{\mu} k^{\nu} ,
\label{eq:1hinv}
\end{equation}
where the amplitudes $A_i$ are functions of all the scalar combinations of
the independent invariants of the process. In particular, $A_4=0$. The 
function $\Delta (k,P_h; S_h=0)$ must meet the applicable constraints 
dictated by hermiticity of fields and invariance under time-reversal 
operations. Hermiticity implies that all $A_i$ amplitudes be real. Since 
the outgoing hadron is described by a free Dirac spinor, time-reversal 
invariance also implies $A_4^* = -A_4$. Combining the two contraints gives 
$A_4 = 0$ in agreement with the previous result deduced just by simple 
algebra arguments. Therefore, no ``T-odd'' structure is generated. This 
result holds true if the hadron plane wave is extended in the complex 
plane~\cite{eikonal} to simulate the effect of an uniform external
potential. The hadron wave function becomes a plane wave damped uniformly 
in space through the imaginary part of $P_h$, but this is not enough to 
generate ``T-odd'' structures in the scattering amplitude. 

Let's now allow for FSI to proceed through a competing channel having a 
different spinor structure with respect to the free channel. As a simple 
test case, we assume for the final hadron spinor the replacement $u(P_h) 
\leftrightarrow u(P_h) + {\rm e}^{{\rm i} \phi} \kslash u(P_h)$, where 
$\phi$ is the relative phase between the two channels. Inserting this back 
into Eq. (\ref{eq:1hsimple}) modifies the $\Delta (k,P_h; S_h=0)$ function 
according to 
\begin{eqnarray}
\Delta (k,P_h; S_h=0) &= &\left[ A_1 (k^2 +1) + B_1 \cos \phi \right] M_h 
+ A_2 (1 - k^2) \Pslash_h + \nn\\
&&\left[ A_3 + B_3 + B_3' \cos \phi \right] \kslash + 
\displaystyle{\frac{B_4 \sin \phi}{M_h}} \sigma_{\mu \nu} P_h^{\mu} 
k^{\nu} \; ,
\label{eq:1hinvdw}
\end{eqnarray}
where the new amplitudes $B_i$ are still scalar combinations of the
invariants of the process. In particular, the coefficient of the tensor 
structure $\sigma_{\mu \nu}$, $B_4=2 M_h/(k^2 - m^2)$, is now not 
vanishing provided that the interference between the two channels, namely 
the phase $\phi$, is not vanishing. In this case, a ``T-odd'' contribution 
arises and is maximum for $\phi = \pi /2$. 

These simple arguments show that, in order to model ``T-odd'' FF in 
one-hadron semi-inclusive processes without giving up factorization, one 
needs to relate the modifications of the hadron wave function to a 
realistic microscopic description of the fragmenting jet. Such a hard task 
suggests that a more convenient way to model occurence and properties of 
``T-odd'' FF is to look at residual interactions between two hadrons in 
the same jet, considering the latter as a spectator and summing over all 
its possible configurations.

\section{Quark-quark correlation function}
\label{sec:three}

In analogy with semi-inclusive hard processes involving one detected 
hadron in the final state~\cite{piet96}, the simplest matrix element for 
the hadronisation into two hadrons is the quark-quark correlation function 
describing the decay of a quark with momentum $k$ into two hadrons 
$P_1, P_2$, namely  
\begin{equation}
\Delta_{ij}(k;P_1,P_2)= \displaystyle{\sum_X} \int
\frac{d^4\zeta}{(2\pi)^4} \; 
e^{ik\cdot\zeta}\langle 0|\psi_i(\zeta)\,a_{P_2}^\dg\,a_{P_1}^\dg |X
\rangle \; \langle X| a_{P_1}\,a_{P_2}\,\ol{\psi}_j(0)|0\rangle \;,
\label{eq:defDelta}
\end{equation}
where the sum runs over all the possible intermediate states involving the 
two final hadrons $P_1,P_2$. Since the three external momenta $k,P_1,P_2$ 
cannot all be collinear at 
the same time, we choose for convenience the frame where the total pair 
momentum $P_h=P_1+P_2$ has no transverse component. The constraint to 
reproduce on-shell hadrons with fixed mass $(P_1^2=M_1^2, P_2^2=M_2^2)$ 
reduces to seven the number of independent degrees of freedom. By defining
the light-cone components of a vector $a$ as 
$a^\mu=\left[a^-,a^+, {\bf a}_\sT\right]$ with 
$a^\pm=\left(a^0\pm a^3\right)/\sqrt{2}$, the independent variables can 
conveniently be reexpressed in terms of the light-cone component of the 
hadron pair momentum, $P_h^-$, of the light-cone fraction of the quark 
momentum carried by the hadron pair, $z_h=P_h^-/k^-=z_1+z_2$, of the 
fraction of hadron pair momentum carried by each individual hadron, 
$\xi=z_1/z_h=1-z_2/z_h$, and of the four independent invariants that can 
be formed by means of the momenta $k,P_1,P_2$ at fixed masses $M_1,M_2$, 
i.e.
\begin{eqnarray}
\tau_h &=& k^2 \quad ; \quad \sig_h = 2k\cdot P_h =\left\{\frac{M_1^2+
{\bf P}_T^2}{z_h\,\xi}+\frac{M_2^2+{\bf P}_T^2}{z_h\,(1-\xi)}\right\}+
z_h\,(\tau_h+{\bf k}_T^2)
\nn \\
\sig_d &=& 2k\cdot (P_1-P_2) =\left\{\frac{M_1^2+{\bf P}_T^2}{z_h\,\xi}
-\frac{M_2^2+{\bf P}_T^2}{z_h\,(1-\xi)}\right\}+z_h(2\xi-1)(\tau_h+
{\bf k}_T^2)-4\,{\bf k}_T\cdot{\bf P}_T \nn \\
M_h^2 &=& P_h^2 = 2\,P_h^+\,P_h^-=\left\{\frac{M_1^2+{\bf P}_T^2}{\xi}
+\frac{M_2^2+{\bf P}_T^2}{1-\xi}\right\} \; , 
\label{eq:invariants}
\end{eqnarray}
with ${\bf P}_T^2=\xi\,(1-\xi)\,M_h^2-(1-\xi)\,M_1^2-\xi\,M_2^2$.

\begin{figure}[h]
\begin{center}
\psfig{file=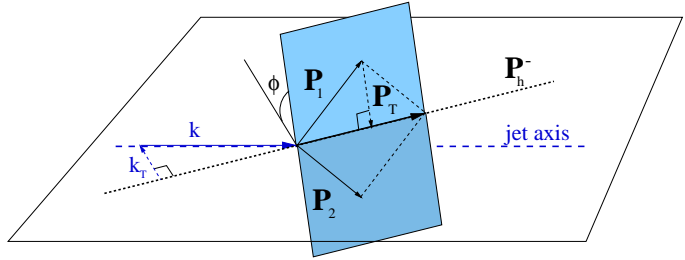, width=9cm}
\end{center}
\caption{Kinematics for a fragmenting quark jet containing a pair of 
leading hadrons.}
\label{fig2}
\end{figure}

By generalizing the Collins-Soper light-cone formalism~\cite{colsop} for 
fragmentation into multiple hadrons, the cross section for 
two-hadron semi-inclusive emission can be expressed in terms of specific 
Dirac projections of $\Delta$ 
after integrating over the (hard-scale suppressed) light-cone component 
$k^+$ and, consequently, taking $\zeta$ as light-like.
Expressing the integrations in a covariant way~\cite{piet96}, we get 
\begin{equation} 
\Delta^{[\Gamma]}(z_h,\xi,P_h^-,M_h^2,\sig_d)=
\int d\sig_h \, d\tau_h \;\delta\left(\tau_h+{\bf k}_\sT^2-
\frac{\sig_h}{z_h}+\frac{M_h^2}{z_h^2}\right)\;\frac{\mbox{Tr}[
\Delta \; \Gamma]}{8z_hP_h^-} \;.
\label{eq:projDelta2}
\end{equation}
Using Eq. (\ref{eq:invariants}) it is possible to reexpress 
$\Delta^{[\Gamma]}$ as a function of $z_h,\xi,{\bf k}_\sT^2,{\bf P}_\sT^2$,
${\bf k}_\sT \cdot {\bf P}_\sT$, where ${\bf P}_\sT$ is 
(half of) the transverse momentum between the two hadrons in the 
considered frame. In this manner $\Delta^{[\Gamma]}$ depends on how much 
of the fragmenting quark momentum is carried by the hadron pair $(z_h)$, 
on the way this momentum is shared inside the pair $(\xi)$, and on the
``geometry'' of the pair, namely on the relative momentum of the two 
hadrons $({\bf P}_\sT^2)$ and on the relative orientation between the 
pair plane and the quark jet axis $({\bf k}_\sT^2,{\bf k}_\sT \cdot 
{\bf P}_\sT$, see also Fig.~\ref{fig2}).

\section{Analysis of interference fragmentation functions}
\label{sec:four}

If the polarizations of the two final hadrons are not observed, the 
quark-quark correlation $\Delta(k;P_1,P_2)$ of Eq. (\ref{eq:defDelta}) can 
be generally expanded, according to hermiticity and parity invariance, as 
a linear combination of the independent Dirac structures of the process 
\begin{eqnarray}
&{}&\begin{array}{l} \Delta(k;P_1,P_2) = C_1\,\left(M_1+M_2\right) + 
    C_2\,\Pslash_1 + C_3\,\Pslash_2 + C_4\,\kslash + 
    \frac{C_{5}}{M_1}\,\sig^{\mu \nu} P_{1\mu} k_\nu + \\
    \begin{array}{l} \frac{C_{6}}{M_2}\,\sig^{\mu \nu} P_{2\mu} k_\nu
    + \frac{C_{7}}{M_1+M_2}\,\sig^{\mu \nu} P_{1\mu} P_{2\nu} + 
    \frac{C_8}{M_1 M_2}\,\g_5\,\eps^{\mu\nu\rho\sig}
     \g_\mu P_{1\nu} P_{2\rho} k_\sig \;. \end{array} \end{array} 
     \label{eq:ansatz}
\end{eqnarray} 
From the hermiticity of the fields it follows that $C_i^* = C_i, \  
i=1,..,12$, and, if constraints from time-reversal invariance can be 
applied, that $C_i^* = C_i, \  i=1,..,4;\  C_i^* = -C_i,\  i=5,..,8$.  
It follows that $C_5=C_6=C_7=C_8=0$, i.e. terms involving
$C_5,..,C_8$ are naive ``T-odd''. Inserting the ansatz~(\ref{eq:ansatz}) 
in Eq.~(\ref{eq:projDelta2}), we get the following Dirac projections
\begin{eqnarray} 
\Delta^{[\g^-]}(z_h,\xi,{\bf k}_\sT^2,{\bf P}_\sT^2,
{\bf k}_\sT \cdot {\bf P}_\sT) &\equiv& D_1 \equiv \int [d\sig_h d\tau_h] 
\  f_{D_1}(C_2,C_3,C_4) \nn \\
\Delta^{[\g^- \g_5]}(z_h,\xi,{\bf k}_\sT^2,{\bf P}_\sT^2,{\bf k}_\sT \cdot 
   {\bf P}_\sT) &\equiv& 
     \frac{\eps_\sT^{ij} \,P_{Ti}\,k_{Tj}}{M_1\,M_2}\; G_1^\perp \equiv 
     \int [d\sig_h d\tau_h ] \  f_{G_1^{\perp}} (C_8) \nn \\
\Delta^{[i\sig^{i-} \g_5]}(z_h,\xi,{\bf k}_\sT^2,{\bf P}_\sT^2,
{\bf k}_\sT \cdot {\bf P}_\sT) &\equiv& {\epsilon_\sT^{ij}P_{Tj}\over 
M_1+M_2}\, H_1^{\newangle}+{\epsilon_\sT^{ij}k_{Tj}\over M_1+M_2}\,
 H_1^\perp \nn\\
\hfil  \equiv \int [d\sig_h d\tau_h] & &\left[ 
  f_{H_1^{\newangle}} (C_5,C_6,C_7) + f_{H_1^{\perp}} (C_5,C_6) \right] 
  \; , \label{eq:proj} 
\end{eqnarray}  
where $\epsilon_\sT^{\mu\nu} = \epsilon^{-+\mu\nu}$ and the integration is
made covariant as in Eq. (\ref{eq:projDelta2}). The functions 
$D_1,G_1^\perp,H_1^{\newangle},H_1^\perp$ are the interference FF that 
arise at leading order for the fragmentation of a current quark into two 
unpolarized hadrons inside the same jet. 
\begin{figure}[h]
\begin{center}
\psfig{file=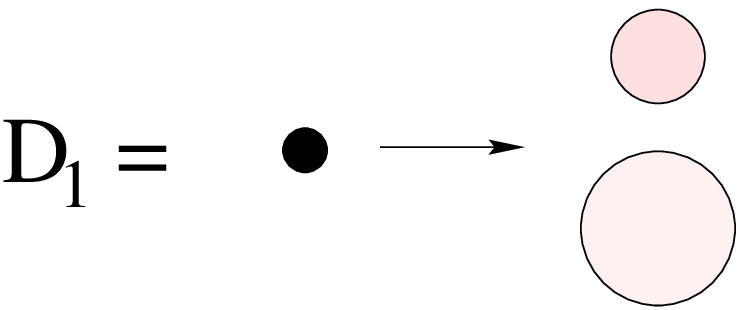, width=3cm} \hspace{.5cm}
\psfig{file=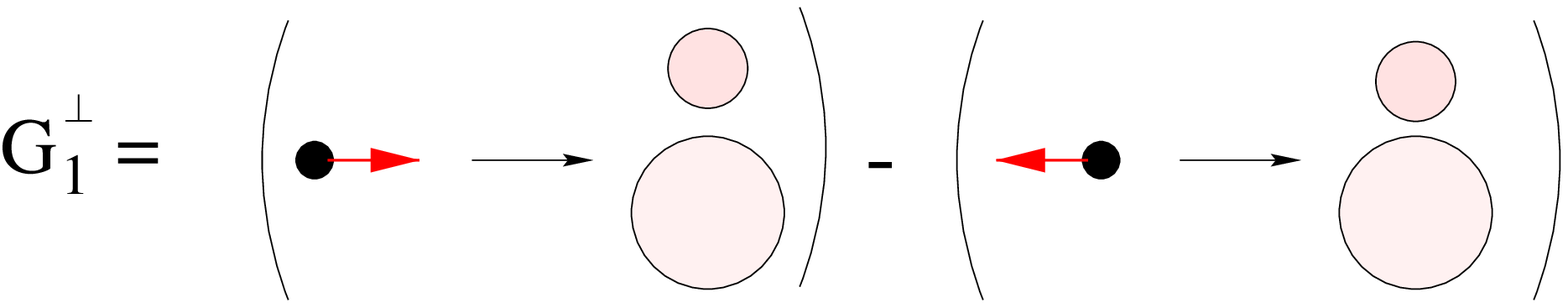, width=7cm} \\
\vspace{.5cm}
\psfig{file=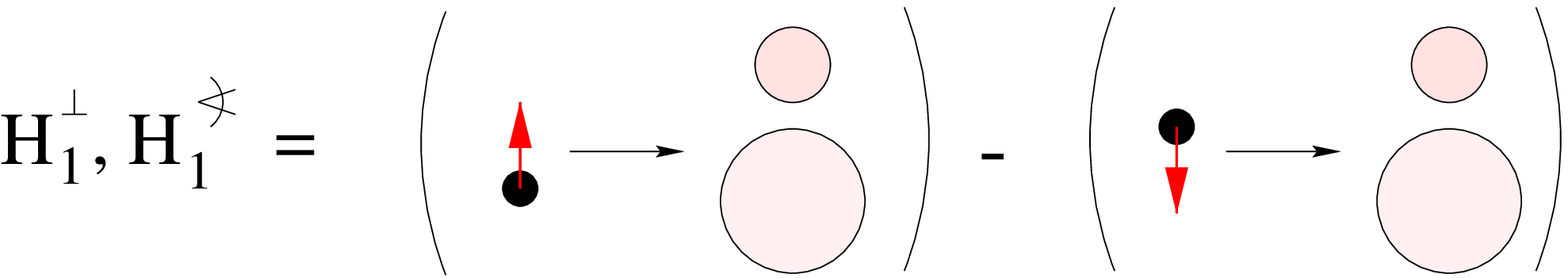, width=7cm} 
\end{center}
\caption{Probabilistic interpretation for the leading order FF arising in 
the decay of a current quark into a pair of unpolarized hadrons.}
\label{fig3}
\end{figure}
The different Dirac structures used in the projections are related to 
different spin states of the fragmenting quark and lead to the nice 
probabilistic interpretation illustrated in Fig.~\ref{fig3}: $D_1$ is the 
probability for an unpolarized quark to produce a pair of unpolarized 
hadrons; $G_1^\perp$ is the difference of probabilities for a 
longitudinally polarized quark with opposite chiralities to produce a pair 
of unpolarized hadrons; $H_1^{\newangle}$ and $H_1^\perp$ both are 
differences of probabilities for a transversely polarized quark with 
opposite spins to produce a pair of unpolarized hadrons. $G_1^\perp$, 
$H_1^{\newangle}$ and $H_1^\perp$ are (naive) ``T-odd'' and do not vanish 
only if there are residual interactions in the final state. In this case, 
the above constraint from time-reversal invariance cannot be applied. 
$G_1^\perp$ is chiral even; $H_1^{\newangle}$ and $H_1^\perp$ are chiral 
odd and can, therefore, be identified as the chiral partners needed to 
access the transversity $h_1$. Given their probabilistic interpretation, 
they can be considered as a sort of ``double'' Collins effect~\cite{coll}.

\section{Spectator model}
\label{sec:five}

\begin{figure}[hbtp]
\begin{center}
\psfig{file=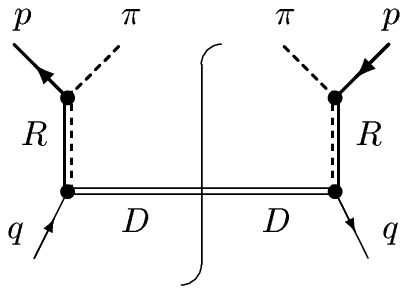, width=3.5cm}\hspace{5mm}
\psfig{file=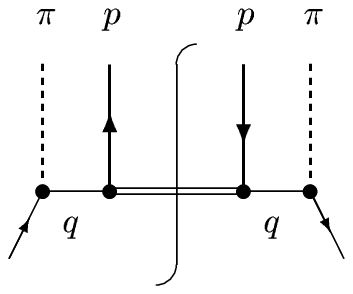, width=2.8cm}\hspace{5mm}
\psfig{file=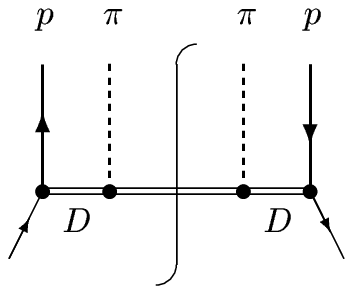, width=3cm}\\
\vspace{2mm}
$a$ \hspace{35mm} $b$ \hspace{35mm} $c$
\end{center}
\caption{\label{fig4}Diagonal diagrams for quark $q$ decay into a proton
$p$ and a pion $\pi$ through a direct channel or a Roper resonance $R$.}
\end{figure}

\begin{figure}[hbtp]
\begin{center}
\psfig{file=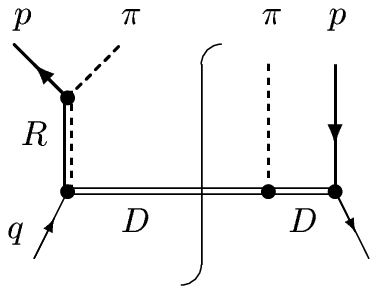, width=3.2cm}\hspace{5mm}
\psfig{file=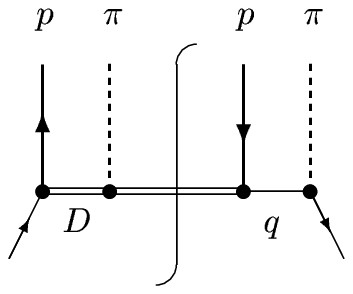, width=2.8cm}\hspace{5mm}
\psfig{file=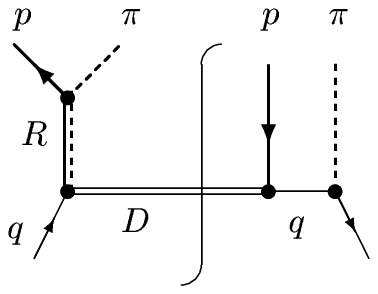, width=3.1cm}\\
\vspace{2mm}
$a$ \hspace{35mm} $b$ \hspace{35mm} $c$
\end{center}
\caption{\label{fig5}Interference diagrams for the same process.}
\end{figure}

So far, the results about the properties of the FF hold true in general
for the quark fragmentation into a pair of unpolarized leading hadrons
at leading order in $1/Q$. In order to make quantitative predictions,
we extend to the present case the formalism of the so-called diquark
spectator model~\cite{spectator}, specializing it to the emission of a
proton-pion pair. The basic idea of the spectator model is to make a 
specific ansatz for the spectral decomposition of the quark correlator by 
replacing the sum over the complete set of intermediate states in 
Eq.~(\ref{eq:defDelta}) with an effective spectator state with a definite 
mass $M_D$, momentum $P_D$ and quantum numbers of the diquark. 
Consequently, the correlator simplifies to 
\begin{equation} 
\Delta_{ij}(k;P_p,P_\pi)\sim
\frac{\theta(P_D^+)}{(2\pi)^3} \; \delta\left((k-P_h)^2-M_D^2\right)\;
\langle 0|\psi_i(0)|\pi,p,D\rangle\langle D,p,\pi|\ol{\psi}_j(0)|0\rangle 
\;,
\label{eq:specDelta}
\end{equation}
where the additional $\delta$ function allows for a completely analytical
calculation of the Dirac projections (\ref{eq:projDelta2}). The quark 
decay is specialized to the set of diagrams 
shown in Figs.~\ref{fig4}, \ref{fig5} and their hermitean conjugates, 
where the interference, necessary to produce the ``T-odd'' FF, takes place 
between the channel for direct production from the quark $q$ of the 
proton-pion pair $(p,\pi)$ and the channel for the decay of the Roper 
resonance $R$.

Assuming that the proton-pion pair has an invariant mass equal to the
Roper resonance one, we can neglect the diagrams not containing the Roper
$R$, such as the \ref{fig4}$b,c$ and \ref{fig5}$b$. Moreover, calculations
are still in a preliminary stage and the results for $G_1^{\perp}$ will be
shown only. This FF is determined mainly by the diagram \ref{fig5}$c$,
while \ref{fig5}$a$ contributes at most to 10\% of the strength. No
contribution comes from the diagonal diagram \ref{fig4}$a$, according to
the ``T-odd'' nature of $G_1^{\perp}$ related to FSI interferences. In 
order to calculate the soft hadronic matrix elements of Eq. 
(\ref{eq:specDelta}) for diagram \ref{fig5}$c$ we assume the most naive 
picture of the quark structure, i.e. in the rest frame all quarks are in 
the $1/2^+$ orbitals and the diquark can be in a spin singlet state 
(scalar diquark, indicated by the label $S$) or in a spin triplet state 
(axial vector diquark, indicated by $A$). Taking the spin-1 field
propagator for the diquark and the Roper propagator quoted by the
PDG~\cite{pdg}, the main Feynman rules for diagram \ref{fig5}$c$ are:
\begin{itemize}
\item $(Rp\pi)$ vertex: $\Upsilon^{Rp\pi}_{ij}=f_{Rp\pi} \;  [\g_5 ]_{ij} 
      \; \equiv g \;  [\g_5 ]_{ij}$ ,\par \noindent
      where $g^2/4 \pi=14.3$ is the strong coupling constant of the 
      $\pi NN$ pseudoscalar interaction. Within the experimental
      uncertainties, the strong Roper coupling can be assumed equal to the
      nucleon one, also because the quark content, and therefore
      the asymptotic form factor, are the same.
      
\vspace{-.3cm}
      
\item $qSR/qSp$ vertex: $\Upsilon^{qSR/qSp}_{ij} = f_{(qSR/qSp)} \; 
      \one_{ij} \equiv N_{qS} \; \displaystyle{
      \frac{\tau_h-m^2}{|\tau_h - \Lambda^2|^2}} \; \one_{ij}$ \par
      \noindent
      
\vspace{-0.3cm}
      
\item $qAR/qAp$ vertex: $\Upsilon_{ij}^{qAR/qAp,\  \mu}=f_{(qAR/qAp)} 
      \; \left[\g_5\g^\mu\right]_{ij} \equiv \displaystyle{
      \frac{N_{qA}}{\sqrt{3}}\; 
      \frac{\tau_h -m^2}{|\tau_h -\Lambda^2|^2}}\; 
      \left[\g_5\g^\mu\right]_{ij}$ \par \noindent
      
\vspace{-.3cm}
      
\item $(q\pi q)$ vertex: $\Upsilon^{q\pi q}_{ij}= f_{q\pi q} \; [ 
      \gamma_5 ]_{ij} \equiv N_{q\pi} \; \displaystyle{
      \frac{\tau_h -m^2}{|\tau_h-\Lambda_{\pi}^2|^{3/2}}} \; 
      [ \gamma_5 ]_{ij}$ 
      
\end{itemize}
\vspace{-.1cm}
\noindent
The introduction of cut-off parameters to exclude large virtualities of
the quark $q$ has been chosen as to kill the pole of the quark 
propagator~\cite{jak97} while keeping the asymptotic behaviour of 
FF at large $z_h$ consistent with the quark counting rule~\cite{joflipat}.
The values themselves of the cut-offs, $\Lambda=0.5$ and 
$\Lambda_{\pi}=0.4$ GeV, as well as the overall normalizations 
$N_{qS}=7.92$ GeV$^2$, $N_{qA}=11.557$ GeV$^2$ and $N_{q\pi}=2.564$ 
GeV$^{1/2}$, are fixed by computing the second moment of $D_1(z_h)$ and 
comparing it with available data; they are taken directly 
from Ref.~\cite{jak97}. 

\section{Numerical Results}
\label{sec:six}

We will plot the $\g^- \g_5$ projection $G_1^{\perp}$ of Eq. 
(\ref{eq:specDelta}) for the fragmentation $u \rightarrow p+\pi$. The 
scalar and axial diquark contributions to the diagram \ref{fig5}$c$ are
combined through the ratio 3:1 to keep the SU(4) structure of the proton
spin-flavor wave function~\cite{jak97}. The parameters take the values 
(in GeV) $m=0.36, M_S=0.6, M_A=0.8, M_h\equiv M_R=1.44,
\Gamma_R=0.35, M_p=0.938, M_{\pi}=0.139$. The
special kinematics ${\bf k}_\sT\cdot {\bf P}_\sT = 0$ is chosen, where the 
hadron-pair plane is perpendicular to the plane containing the jet axis 
and the pair leading light-cone direction $P_h^-$ (see Fig.~\ref{fig2}).
From Eq. (\ref{eq:invariants}) it can be shown that $G_1^{\perp}$ actually
becomes function of $z_h,\xi,{\bf k}_\sT^2$.
\begin{figure}[h] 
\begin{center}
\epsfig{file=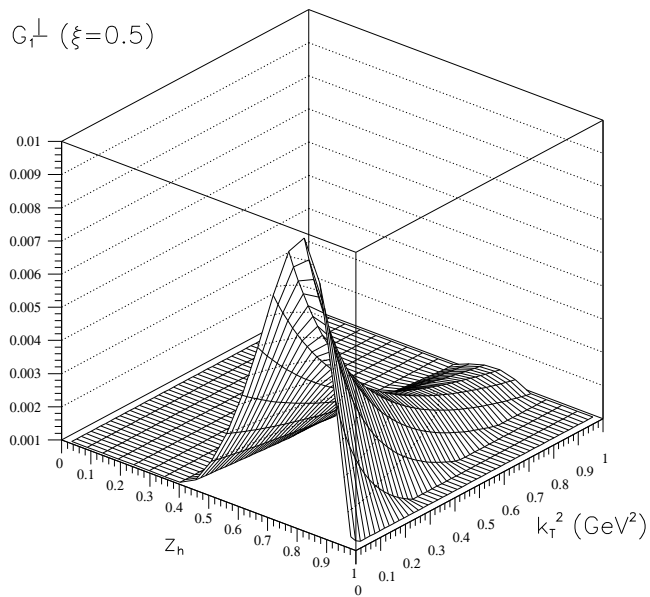, width=9cm}
\end{center}
\caption{\label{fig6}$G_1^{\perp}(z_h,{\bf k}_\sT^2)$ at 
$\xi=0.5$ for the fragmentation of a quark 
$u$ into a proton $p$ and a pion $\pi$. Kinematics is chosen such that the 
invariant mass of the pair is equal to the Roper resonance and 
${\bf k}_\sT\cdot{\bf P}_\sT=0$ (see Fig.~\protect{\ref{fig2}}).}
\end{figure}

In Fig. \ref{fig6} $G_1^{\perp \; u \rightarrow p+\pi}(z_h,{\bf k}_\sT^2)$
is shown for $\xi=0.5$. We have checked that the result is rather
insensitive to $\xi$. On the contrary, the maximum sensitivity 
to the fragmentation mechanism is concentrated around the kinematical 
range where the pair takes roughly 70\% of the jet longitudinal momentum 
and has a small transverse momentum with respect to the jet axis. By 
``cutting'' the 3-d surface at constant values of $z_h\ge 0.6$, one can
obtain curves that, for increasing $z_h$, get concentrated at lower 
${\bf k}_\sT^2$ and have an increasingly less important tail. In other
words, the more the hadron pair is leading, i.e. it takes most of the jet
longitudinal momentum, the more $G_1^{\perp}$ is concentrated around the 
jet axis with smaller transverse momentum. 
\vspace{.5cm}

Work is in progress to complete the calculation of FF at leading order
including also chiral odd $H_1^{\newangle}$ and
$H_1^{\perp}$~\cite{paper}. Possible asymmetry measurements will also be
addressed that allow isolation of each individual FF. In particular, after 
integration over ${\bf k}_\sT$, the combination of $H_1^{\newangle}$ and 
the transversity distribution $h_1$ could be isolated in the cross section 
for a semi-inclusive DIS on a polarized nucleon target, where an asymmetry 
can be built by measuring the proton-pion pair at some angle ${\bf k}_\sT
\cdot {\bf P}_\sT$ with respect to the jet axis and then 
exchanging the mutual position inside the pair, i.e. flipping ${\bf P}_\sT
\rightarrow -{\bf P}_\sT$. This and other possibilities are presently 
under consideration~\cite{paper}.

\section*{Acknowledgments}

This work is part of the TMR program ERB FMRX-CT96-0008.
Interesting and fruitful discussion with P. Mulders are greatly 
acknowledged.

\end{document}